\documentclass[sigconf,authorversion]{acmart}

\AtBeginDocument{%
  \providecommand\BibTeX{{%
    \normalfont B\kern-0.5em{\scshape i\kern-0.25em b}\kern-0.8em\TeX}}}

\setcopyright{acmcopyright}
\copyrightyear{2019}
\acmYear{2019}

\acmConference[MIG'19]{MIG'19}{Nov. 28--30, 2019}{Newcastle, UK}
\acmBooktitle{MIG'19: Motion, Interaction and Games, Nov. 28--30, 2019, Newcastle, UK}

\newcommand{\dr}[1]{\textcolor{black}{#1}}

\setcopyright{acmlicensed}

\begin{document}

\copyrightyear{2019} 
\acmYear{2019} 
\acmConference[MIG '19]{Motion, Interaction and Games}{October 28--30, 2019}{Newcastle upon Tyne, United Kingdom}
\acmBooktitle{Motion, Interaction and Games (MIG '19), October 28--30, 2019, Newcastle upon Tyne, United Kingdom}
\acmPrice{15.00}
\acmDOI{10.1145/3359566.3360067}
\acmISBN{978-1-4503-6994-7/19/10}

\title{Animation Synthesis Triggered by Vocal Mimics}

\author{Adrien Nivaggioli}
\author{Damien Rohmer}
\affiliation{%
  \institution{LIX, \'Ecole Polytechnique/CNRS, IP Paris}
  \country{France}}

\renewcommand{\shortauthors}{A. Nivaggioli, D. Rohmer}

\begin{abstract}
  We propose a method leveraging the naturally time-related expressivity of our voice to control an animation composed of a set of short events. The user records itself mimicking onomatopoeia sounds such as "\emph{Tick}", "\emph{Pop}", or "\emph{Chhh}" which are associated with specific animation events. The recorded soundtrack is automatically analyzed to extract every instant and types of sounds. We finally synthesize an animation where each event type and timing correspond with the soundtrack. In addition to being a natural way to control animation timing, we demonstrate that multiple stories can be efficiently generated by recording different voice sequences. Also, the use of more than one soundtrack allows us to control different characters with overlapping actions.
\end{abstract}



\begin{teaserfigure}
  \includegraphics[width=\textwidth]{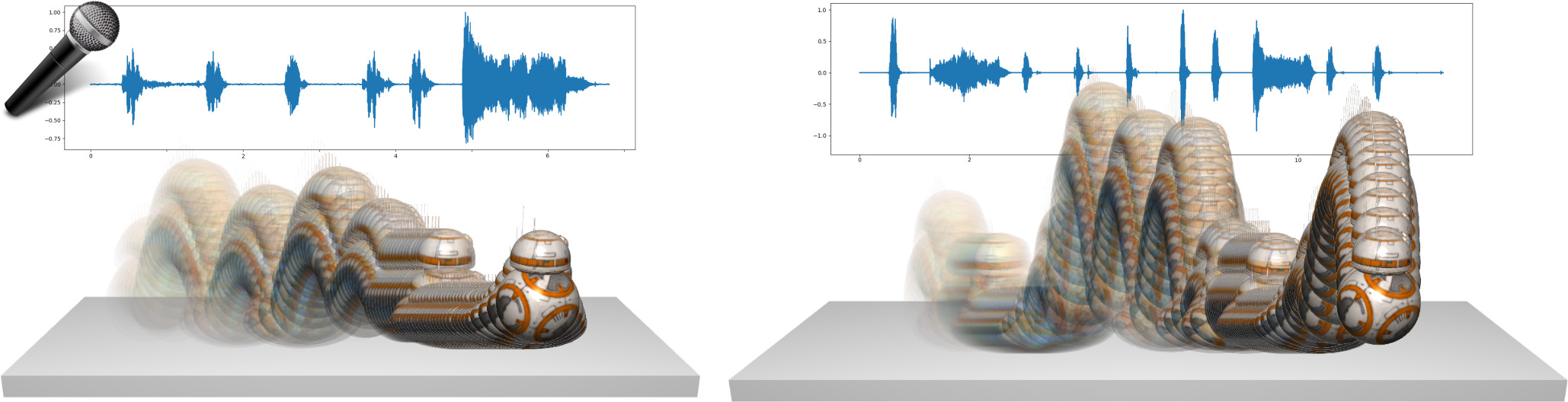}
  \caption{Two different animated scenarios containing jumps and sliding of the virtual character are synthesized from two different vocal sound sequences and procedurally synchronized with them. Each short animated event is associated to the sound "\emph{Tick}", "\emph{Pop}" or "\emph{Chhh}". Our approach automatically detect these specific sounds from the recorded soundtrack and triggers the associated animation.}
  \Description{Teaser}
  \label{fig:teaser}
\end{teaserfigure}

\maketitle

\sloppy

\section{Introduction}

Computer animation generation is an essential tool for entertainment industries such as animation studios or video games developers. With the exception of complex dynamic phenomenons requiring physically based simulations, the fundamental principle of computer animation for virtual characters is mostly based on key-framing, meaning that a model should match a predefined shape at specific times, while in-betweens can be automatically computed using interpolation schemes. Defining these key-times is an important step and is called \emph{timing} by animators.

While shape deformation and interpolation have been widely studied, fewer works targeted the specific case of timing set-up. Standard production softwares used for key-frame animation such as Maya model the time as a 1D axis on which the animator pins some key-poses, and possibly adapts the geometrical parameters using \emph{animation-curves}, i.e. 2D curves expressing the evolution of each degree of freedom with respect to time.
While such approach allows a global and precise view of the entire timing of the animation, the time dimension is represented as a spatial one. A constant mental conversion between space and time -- associated with intellectual effort and experience to master -- is therefore required when setting up these animations.

In this work, we propose to take advantage of the natural time-related expressivity of our voice to control animation timing without the use of any manual space-time curve definition. 
More precisely, a user records a sequence containing different expressive sounds such as onomatopoeias ("\emph{Boum}", "\emph{Zap}", "\emph{Bang}", etc), acting as triggers for basic actions constituting the entire animation. Thus, the timing of the recorded sound sequence defines the timing of the animation. This sound sequence is automatically analyzed to extract every individual expressive sound-related event, and the correspondence between a specific sound and the corresponding action is encoded in a predefined database. As a result, the user-made soundtrack can be efficiently converted into a set of different keyframes, defined and parameterized by the type of sound, while the timing is naturally encoded by the time-position in the soundtrack. The final animation can finally be run using some procedural animation or interpolation.

While our approach doesn't target the fine authoring reached by standard animation-curves setup, it allows for a very fast and expressive setup of rough animation. Indeed, mimicking sound is a very natural and intuitive way to convey the sense of both timing and action. Pre-productions of movie animation could benefit from such approach to efficiently generate animated storyboards, depicting only rough graphics, but already including the temporal information of the main actions. Games industry could also find applications where a player could mimic with his voice some actions, and see the result of his voice record as an automatically adapting story. Interestingly, some games controlled by voice sounds are already available (Yasuhati~\shortcite{Yasuhati}, Chicken Scream~\shortcite{ChickenScream}, Karaoke Revolution~\shortcite{KaraokeRevolution}), but the controls are restricted so far to the magnitude or pitch of the sound, without allowing explicit authoring on the type of event.

\section{State of the art}

Generating animation and sound is a long-standing problem in the Computer Graphics community.
While the efficient and realistic generation of sound from an existing 3D scene remains a highly studied research topics~\cite{Takala1992,Doel-FoleyAutomatic-SIGGRAPH-2001,Wang2018}, we focus in this work on the converse, namely, the synthesis of an animation given its sound as input.

An extensive research literature can be found on human face synthesis from speech inputs used for virtual agent animation~\cite{Pelachaud1996, vougioukas2019realistic} and goes beyond the scope of our approach. 
In the most recent works, both speech recognition and expressive facial motion synthesis take advantage of deep-learning approaches thanks to large available databases~\cite{Zhou-Visemenet-TOG-2018,Pham-ICMI-2018,Cudeiro-VOCA-CVPR-2019}. Retargeting real video even achieve striking realistic results hard to differentiate from real records~\cite{Obama2017,Vougioukas-CVPR-2019}. 
Pure speech-based command has also been explored in several commercial entertainment products related, for instance, to video games (Mass Effects 3, Nevermind, etc.), VR (Oculus Voice, Alexa Skills Kit, etc.), and personal assistants (Google Assistant, Alexa, etc.). \\
While achieving very high quality results, the complex pipeline used for speech analysis is very specific and doesn't fit directly to our objective. 
Indeed, human speech is highly structured, and usually analyzed through unit sound components called phonemes. Speech recognition thus focus on the robust extraction of, possibly complex, words and sentences from a common static dictionary defined by a given language. In our case, we rather aim at extracting quite simple onomatopoeia sounds from a single user, but these sounds are expected to be easily changed and adapted to the animated scene. Moreover, the mapping between the sound and the animated sequence is also scene dependent. This approach doesn't fit well to data-based learning approaches, and we therefore rely on more standard signal processing tools to analyze a generic sound signal for which the mapping between the sound pattern and the animation is directly provided by the user.

Fuzzy mapping between sound signals and animation has also been studied in the case of musical inputs. Extracting musical characteristics, and in particular its notion of rhythm and beats, led to methods able to edit motions~\cite{CARDLE_2002}, fully synthesize a dancing characters~\cite{Kim-SIGGRAPH-2003, Shiratori-DancingToMusic-EG-2006,Sauer-MusicDrivenCharacterAnimation-2009}, visualize the music by mapping the mesh harmonics to the sound frequencies~\cite{lewiner2010tuning}, or animate the musician~\cite{Shlizerman-AudioToBodyDynamics-CVPR-2018}. Given an existing background music and an animation, pure synchronization can be handled by adapting the timing of both inputs~\cite{Lee-Synchronization-EG-2005}. Finally, video synchronization with a given music was also explored~\cite{Liao-VideoMontage-SIGGRAPH-2015,Davis-VisualBeat-2018}.

Closer to our objectives, Langlois and James~\shortcite{Langlois-InverseFoley-SIGGRAPH-2014} proposed a method to synthesize generic rigid-body motions, and in particular use sound to trigger impacts and frictions. They fully automatize the animation generation for the specific case of rigid bodies, but the sound analysis is performed manually as the user is asked to pin key-times corresponding to contact sounds. On the opposite, our approach automatizes the sound extraction and mapping to generic type of events, but requires the user to script the associated triggered animation.

In the following, we explain in Chap.~\ref{chap:sound_setup} the sound sequence setup and analysis to extract specific patterns. Then we detail in Chap.~\ref{chap:scene} the usage of such soundtrack in the case of different animated scenario, before discussing limitations and possible extensions in Chap.~\ref{chap:conclusion}.

\section{Analyzing the sound sequence to extract animation related patterns}
\label{chap:sound_setup}

Let us consider a set of possible \emph{types} of events of an animated scene such as a ball touching and bouncing on the floor, hitting a wall or sliding on a surface. These possible events are called event-\emph{types}.
Note that we consider two different categories of event-types: \emph{impulse} events, modeling an instantaneous change of state (ex. ball hitting the floor triggering a bouncing animation), and \emph{continuous} ones, for which the corresponding action may last for some time (ex. sliding).

In a preparatory stage, each type of event \(i\) is manually associated with a specific onomatopoeia sounds, similar to a word in a dictionary of events. We call \emph{sound pattern} \(p_i\) the short generic sound signal associated to a given type of event.
Once the association between events and sound patterns is made, the user records a sound sequence containing several onomatopoeia sounds played in arbitrary order and time. We call \(s\) the signal associated to this sequence, and suppose conventionally that \(s(t)\in[-1,1]\). Next, this sound sequence is automatically processed to extract each sound pattern and their time-positions.

\begin{figure}[!ht]
  \includegraphics[width = \linewidth]{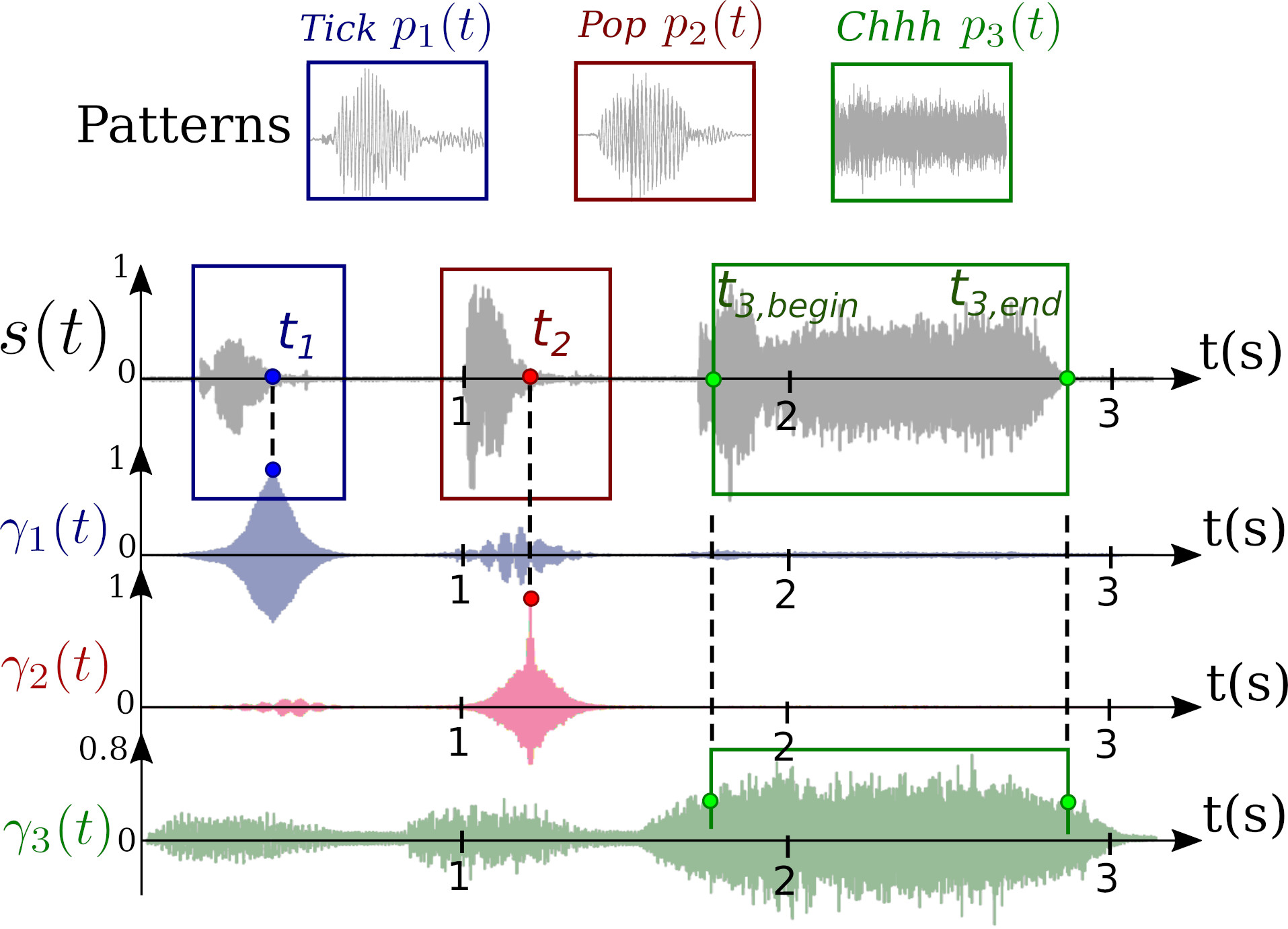}
  \Description{Extraction of a sound pattern}
    \caption{Extraction of instances for three different sound patterns. \(s(t)\) is the recorded sound sequence, and the three following lines: \(\gamma_1(t)\), \(\gamma_2(t)\), \(\gamma_3(t)\) are the cross-correlation between the signal and the respective pattern. Their local maxima indicate the time where each event should be triggered.}
    \label{fig:detection}
\end{figure}

Variability is an inherent part of sounds generated by human voices, therefore the full sequence s will never contain an exact copy of any sound pattern \(p_i\), but slight variations of it that we call \emph{instances} of the sound pattern. Our detection relies on robustly finding similarities between two signals \(s\) and \(p_i\) using the cross-correlation signal \(\gamma_i\) \dr{used as a simple and versatile readily available tool}, where
\[\gamma_i(t) = \int_{u\in\mathbb{R}} s(u)\;p_i(u+t)\,\mathrm{d}u \mbox{ .}\]
The local maxima of \(\gamma_i\) indicates the precise localization of each instance of the sound type \(i\). Therefore, computing all cross-correlations between \(s\) and all sound patterns allow us to extract their respective local maxima and retrieve all instances of the events. An example of such detection using three sound patterns "\emph{Tick}" (impulse event), "\emph{Poc}" (impulse event), and "\emph{Chhh}" (continuous event) is illustrated in Figure~\ref{fig:detection}.

Note that \(\gamma_i\) may have a noisy shape and different -- but sound alike -- patterns \(p_i\) and \(p_j\) (ex. "\emph{Boum}", and "\emph{Poum}")  which may both be associated with local maximal values of their respective cross-correlation \(\gamma_i,\gamma_j\) in the same short time period. To avoid ambiguities, we consider a sound-event of type \(i\) to take place at time \(t_k\) only if its cross-correlation have a large enough value \(\gamma_i(t_k)>0.5\), and such that no greater local maxima of any other cross-correlation take place in the same period of time defined by \([t_k-\Delta_t/2,\;t_k+\Delta_t/2]\), where \(\Delta_t\) is the time length of the sound pattern.

In the case of an impulse event, we store for each instance the single time \(t_k\) in which the sound is played. In the case of continuous event, the sound may last for an arbitrary amount of time and cannot be associated with a single time instant. We consider the continuous period of time such that the cross-correlation remains, in average, above a limit threshold to define the beginning and end times \([t_{k,begin},\; t_{k,end}]\) on which the instance is played.

Once an instance of a sound pattern is detected in the sequence, we can compute the \emph{strength} of the current instance to add an extra expressive parameter that can be used to tune the animation sequence associated with the event. 
\dr{Let us call \(x\), subset of \(s\), the signal corresponding to an instance of the sound pattern \(p_i\) detected at time \(t_k\) such that
\begin{equation*}
  x(t)= 
  \left\{
    \begin{array}{ll}
      s(t)\,,& 
      \left|
      \begin{array}{l}
        t\in [t_k-\Delta_t/2, t_k+\Delta_t/2] \;  \mbox{for an impulse event} \\
        t\in [t_{k,begin},t_{k,end}] \;\; \mbox{for a continuous event} \\
      \end{array}
      \right. \\
      0\;,& \mbox{otherwise.}
    \end{array}
  \right.
\end{equation*}
}
The sound \emph{strength} is related to the energy of the associated signal \(\mathcal{E}\) computed as
\[\mathcal{E}(x)=\int_{t\in\mathbb{R}} x^2(t)\mathrm{d}t \mbox{ .}\]
As we are interested in the relative \emph{strength} of \(x\) with respect to the reference given by its pattern \(p_i\), we finally expressed this \emph{strength} by \(E(x) = \left(\mathcal{E}(x)/\mathcal{E}(p_i)\right)^{1/2}\).

\section{Synthesizing 3D animation from sound-related keyframes}
\label{chap:scene}

\emph{Please note that all the animated examples that we describe and their soundtracks are available in the associated video.}

\begin{figure}[!ht]
  \centering
  \includegraphics[width = \linewidth]{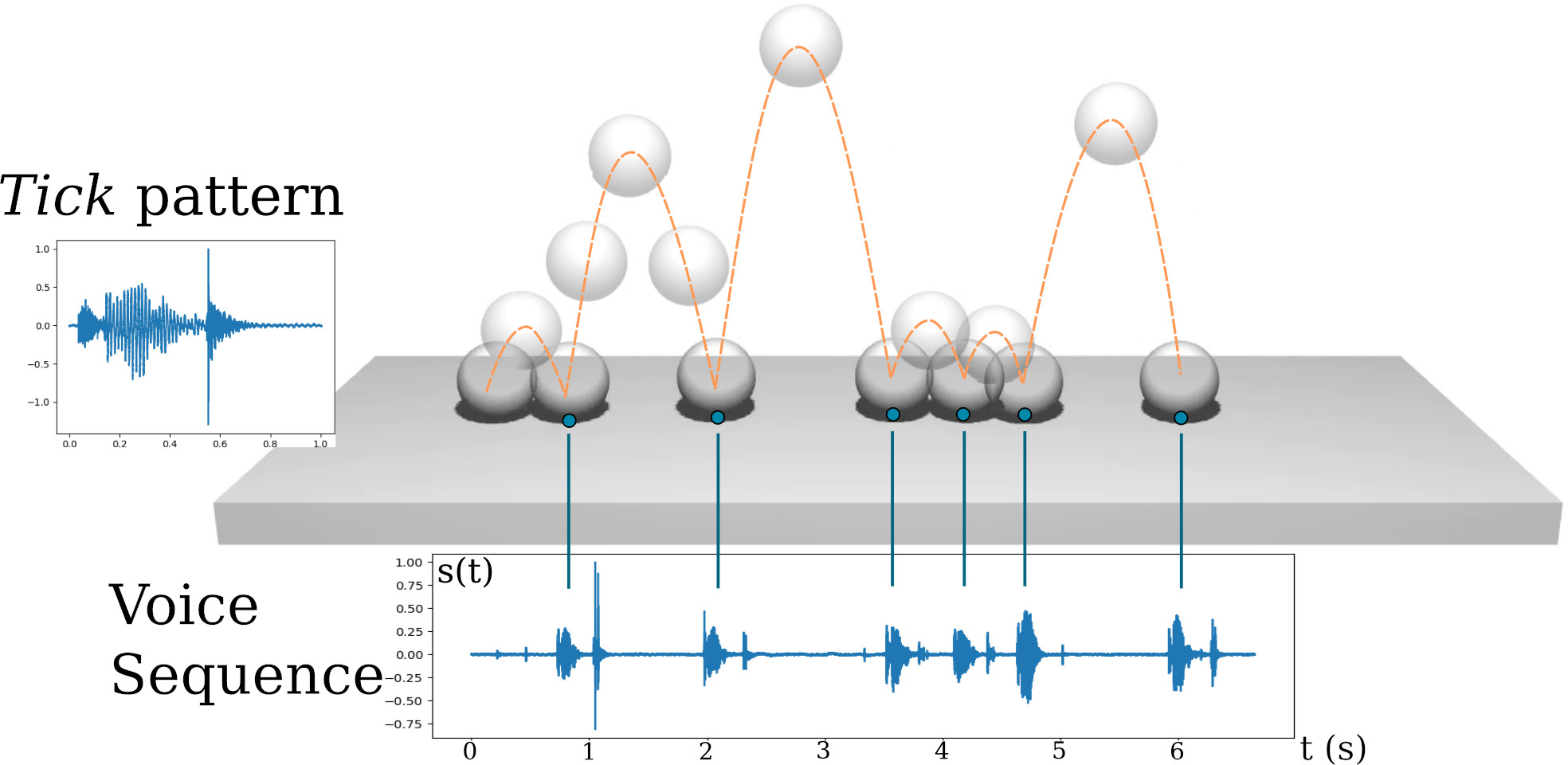}
  \Description{Tick bouncing}
  \caption{Sphere bouncing on the floor at every "\emph{Tick}" instance detected in the recorded voice sequence. Note that the timing of the animation is fully defined by the soundtrack.}
  \label{fig:sphere}
\end{figure}

We first consider a simple scene made of a falling sphere bouncing on the floor. The animation contains only one action triggered by the sound "\emph{Tick}" corresponding to the time where the sphere hits the ground.
The user record a series of "\emph{Tick}" sounds and the associated animation is automatically computed as seen in Figure.~\ref{fig:sphere}. We compute the trajectory \(p(t)=(x(t),y(t),z(t))\) of the sphere using the basic ballistic equation \(p(t)=(0,0,-g)\, t^2/2+v^0\, t+p^0\), where \(g\) is the gravitational constant, \(v^0\) and \(p^0\) are the initial velocity and position.
Every instance "\emph{Tick}" at time \(t_k\) should correspond to the constraint \(z(t_k)=0\). We enforce this constraint in adapting the new bouncing speed in the \(z\) direction after each collision at time \(t_k\) such that \(v_z(t_k)=g\, (t_{k+1}-t_{k})/2\), ensuring that the next floor hit will happen at time \(t_{k+1}\).

\begin{figure}[!ht]
  \centering
  \includegraphics[width = \linewidth]{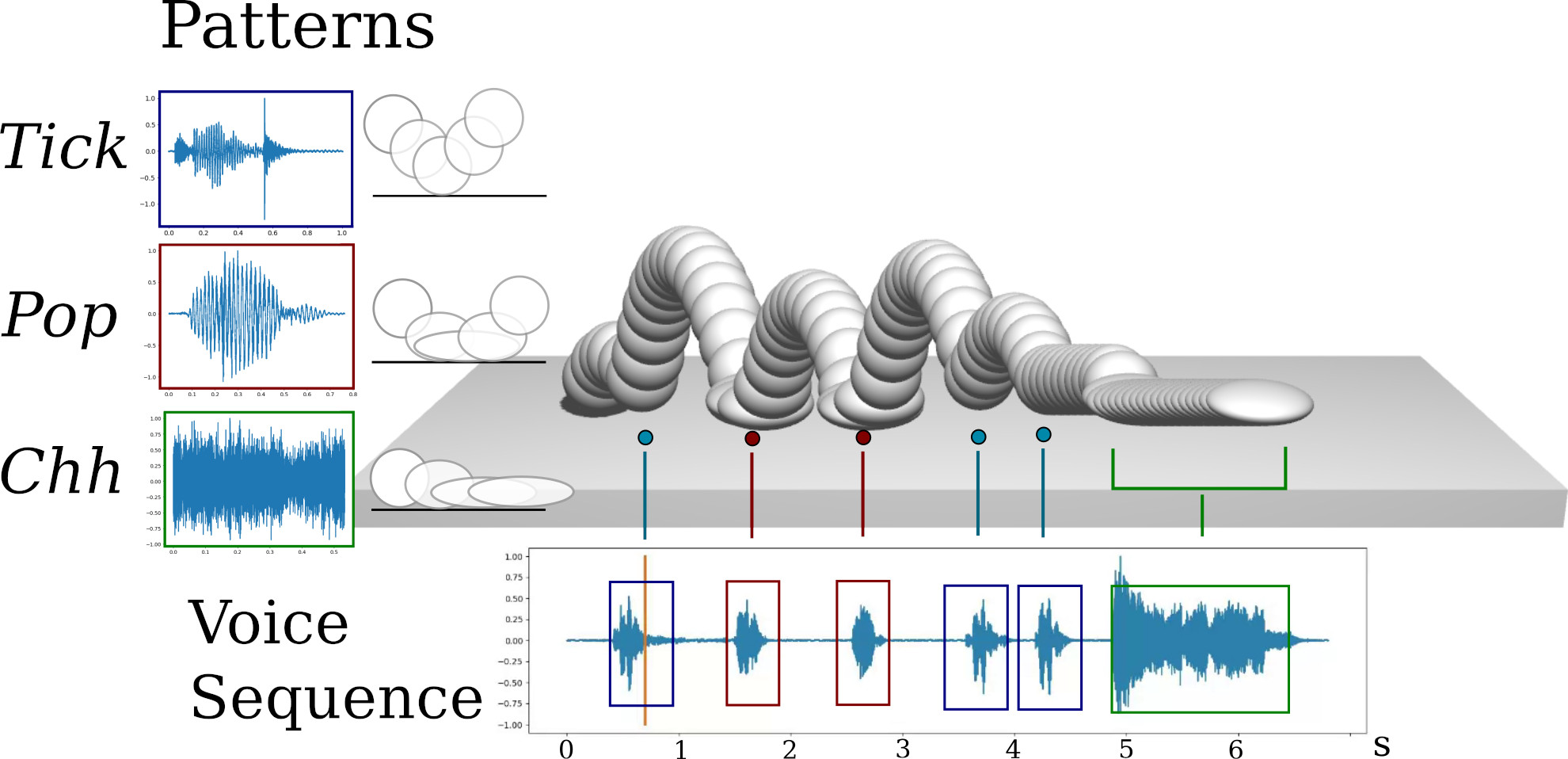}
  \Description{Multiple patterns}
  \caption{Animated sequence using three different sound patterns. Various instances of these patterns are detected in the voice sequence and lead to the three respective events: hard bounce (blue), soft bound (red), and continuous sliding (green).}
  \label{fig:sphere-3-patterns}
\end{figure}

A second scene is generated by taking into account event-types for a bouncing sphere illustrated in Figure~\ref{fig:sphere-3-patterns}: A hard bounce associated to the sound "\emph{Tick}", a soft bounce where the sphere squashes during bouncing impact associated to the sound "\emph{Pop}", and a continuous friction associated to squashing happening as long as the sound "\emph{Chhh}" is made. The instances of the different event-types are detected using the approach based on cross correlation described previously.

More entertaining scenes can be generated in mapping the deformation of the previous sphere model to an arbitrary virtual character squashed uniformly. Figure~\ref{fig:teaser} illustrates two different animated sequences using different soundtracks as inputs applied to a Star-Wars character jumping and sliding on the floor. The two animations use the same parameters, but the change of order and timing of each recorded soundtrack allows us to model different animated stories. Note that the left part of Figure~\ref{fig:teaser} corresponds to the same soundtrack as the one used for the previous ball deformation.

\begin{figure}[!ht]
  \includegraphics[width = \linewidth]{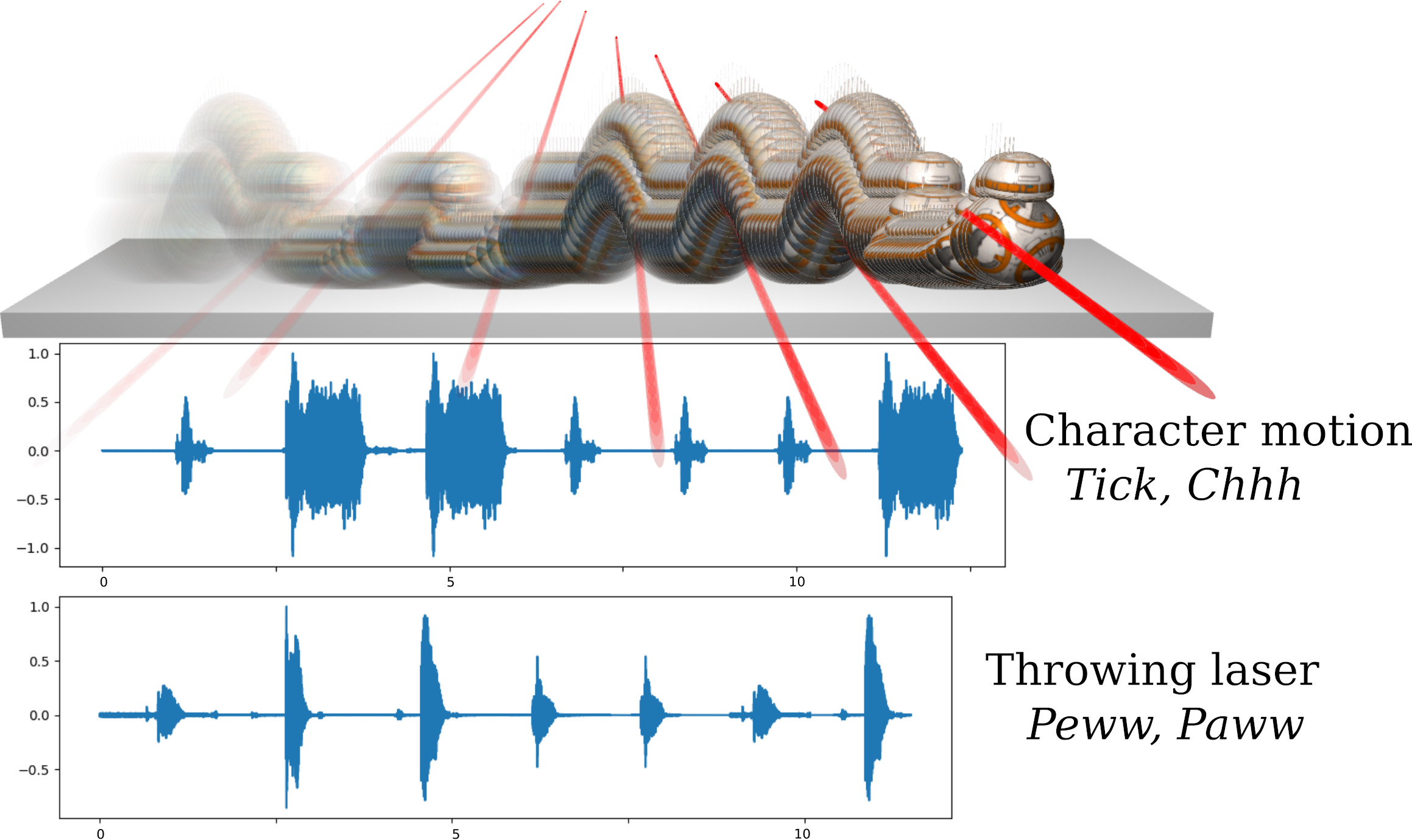}
  \Description{Lasers}
  \caption{Animation containing two elements: a laser and the virtual character avoiding it. A first soundtrack is used to trigger the laser shots, while a second soundtrack is used to trigger the character jumping and sliding. The use of two soundtracks allows us to handle partially overlapping actions and sounds.}
  \label{fig:bb8-laser}
\end{figure}

In the case where the animation contains several overlapping events, multiple soundtracks can be used. Figure~\ref{fig:bb8-laser} shows the case where the virtual character tries to escape some lasers shots. Two types of shots are respectively associated to the sounds "\emph{Peww}" corresponding to low shots, and "\emph{Paww}" corresponding to high shots. Once recorded and visualized, the user may vocalize, and records, a second soundtrack expressing the motion of the character to avoid these laser shots using jumping (with "\emph{Tick}") and sliding (with "\emph{Chhh}"). The final scene shows overlapping events applied to two different virtual models, both synchronized with the soundtrack.

\begin{figure}[!ht]
  \includegraphics[width = \linewidth]{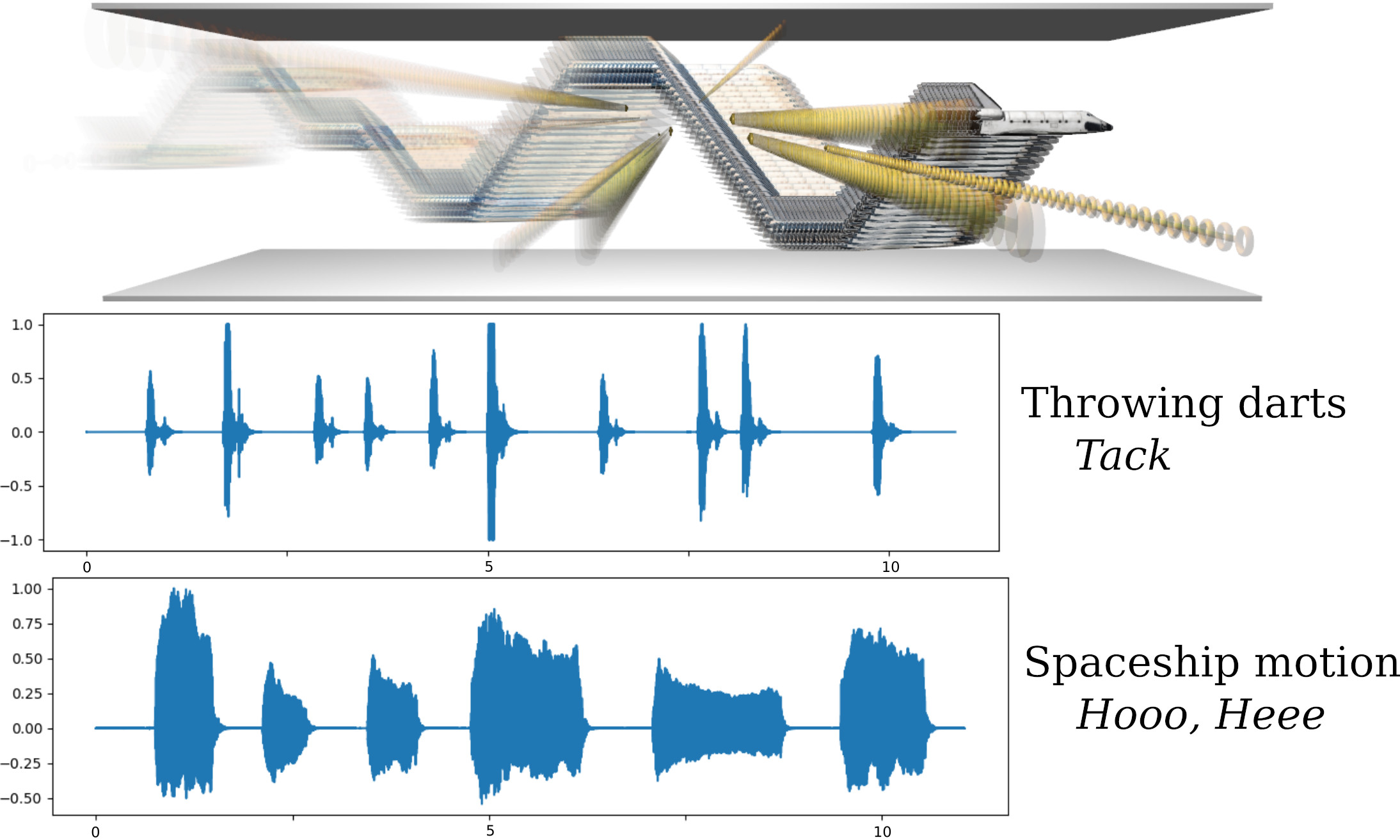}
  \Description{Shuriken example}
  \caption{A game-like animation where a spaceship moves up and down guided by two continuous sound patterns and tries to avoid the yellow darts generated from a second soundtrack.}
  \label{fig:shuriken}
\end{figure}

In the same type of idea, a game-like scene is illustrated in Figure~\ref{fig:shuriken}. In this example, a first recorded sequence of short "\emph{Tack}" sounds is associated to the throw of a dart. The size of each dart is parameterized by the strength of the corresponding "\emph{Tack}" sound. \dr{Note that our example used a linear increase of the dart size with respect to the sound strength \(E(x)\).}
The spaceship can avoid these darts by moving up and down when the respective continuous sounds "\emph{Hooo}" and "\emph{Heee}" are pronounced, and recorded in a second soundtrack. Note that the ascending and descending phase of the spaceship are synchronized with the instant and time length of each continuous sound.

\begin{figure}[!ht]
  \includegraphics[width = \linewidth]{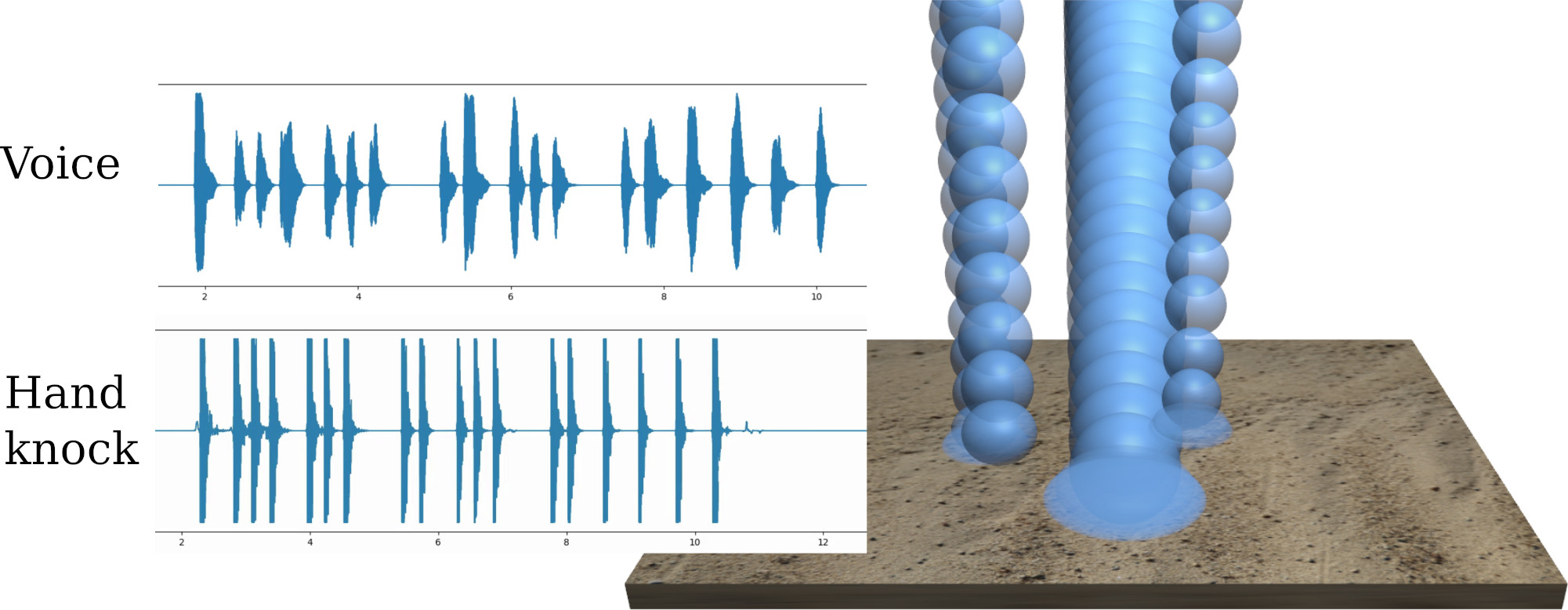}
  \Description{RainDrops}
    \caption{Rain drops scene triggered by impulse sounds following the dynamic of the Tetris music. The first soundtrack corresponds to a vocal record, while the second one corresponds to a hand knocking on a table.}
    \label{fig:rain}
\end{figure}

Finally, another type of scene modeling raindrops is illustrated in Figure~\ref{fig:rain}. Every sound event corresponds to the impact of a rain drop on the floor at a random location. The soundtrack is recorded in two different situations: firstly a vocal record where the sound pattern is given by the sound "\emph{Pom}", and secondly a series of sound impacts made by knocking the hand on a table. In the first case, the voice magnitude of each sound instance parameterizes the size of the drop. In the latter case, the record doesn't correspond to a vocal record but can still be used by our approach given the appropriated sound pattern (the magnitude is not taken into account in this case). In both scenario, the sound events are following the rhythm of Tetris music.

\dr{All the visual animations presented in this work and shown in the associated video are synthesized and rendered on the fly in real time while playing the associated soundtrack in parallel. The preprocess step, and in particular, the computation of the cross-correlation \(\gamma_i\) with a single pattern signal takes around 1 to 3 seconds for an input signal up to \(15s\). The entire computation over all patterns as well as the detection of peaks remains under \(15s\) of computational time for all our examples. Note that we used a standard out-of-the box correlation implementation -- NumPy package in Python -- and applied over the entire input signal without any optimization.}

\section{Conclusion and Future Works}
\label{chap:conclusion}

We proposed a method to generate animated scenarios composed of short events from a vocal -- or hand-made -- soundtrack. Our approach offers an intuitive control of the timing of the animation through the natural time-expressivity of sounds sequence. Moreover, multiple version of a scenario can efficiently be synthesized by recording different soundtracks.

This work leads to several avenues for improvements. Our sound extraction is currently using simple tools based on cross-correlation on the entire signal. The approach may lack robustness for close-by sounds that can be miss-detected, and requires the entire soundtrack to be pre-recorded before looking at the result. More advanced and efficient sound analysis can both increase its robustness and speed, and developing a real-time visualization of our vocal input could be a future extension well adapted to video games for instance. Sound parameters such as pitch could also be used to further parameterize the visual animation, and combining our approach with existing speech recognition tools could be used to offer a higher level of control.

Another possible extension relates to the mapping between the sound pattern and the corresponding animation which is fully predefined by manual scripting in this work. A more automatic approach allowing us to match existing animation characteristics with sound signal~\cite{Cardle-MotionDriven-SCA-2003} could ease this process.

\bibliographystyle{ACM-Reference-Format}
\bibliography{bibliography}

\end{document}